\documentclass[american,english,aps,manuscript]{revtex4}
\usepackage[T1]{fontenc}
\usepackage[latin9]{inputenc}
\usepackage{units}
\usepackage{amstext}
\usepackage{graphicx}

\makeatletter

\providecommand{\tabularnewline}{\\}
\newcommand{\lyxdot}{.}

\@ifundefined{textcolor}{}
{%
 \definecolor{BLACK}{gray}{0}
 \definecolor{WHITE}{gray}{1}
 \definecolor{RED}{rgb}{1,0,0}
 \definecolor{GREEN}{rgb}{0,1,0}
 \definecolor{BLUE}{rgb}{0,0,1}
 \definecolor{CYAN}{cmyk}{1,0,0,0}
 \definecolor{MAGENTA}{cmyk}{0,1,0,0}
 \definecolor{YELLOW}{cmyk}{0,0,1,0}
}

\makeatother

\usepackage{babel}
\begin{document}

\title{Potential energy curves for P$_{2}$ and P$_{2}^{+}$ constructed
from a strictly N-representable natural orbital functional}

\author{M. Piris$^{1,2,3}$, N. H. March$^{1,4,5}$\bigskip{}
}

\address{$^{1}$Donostia International Physics Center (DIPC), 20018 Donostia,
Spain.}

\address{$^{2}$Kimika Fakultatea, Euskal Herriko Unibertsitatea (UPV/EHU),
20018 Donostia, Spain.}

\address{$^{3}$IKERBASQUE, Basque Foundation for Science, 48011 Bilbao, Spain.}

\address{$^{4}$Department of Physics, University of Antwerp, Antwerp, Belgium.}

\address{$^{5}$Oxford University, Oxford, England.\bigskip{}
}
\begin{abstract}
The potential energy curves of P$_{2}$ and P$_{2}^{+}$ have been
calculated using an approximate, albeit strictly N-representable,
energy functional of the one-particle reduced density matrix: PNOF5.
Quite satisfactory accord is found for the equilibrium bond lengths
and dissociation energies for both species. The predicted vertical
ionization energy for P$_{2}$ by means of the extended Koopmans'
theorem is 10.57 eV in good agreement with the experimental data.
Comparison of the vibrational energies and anharmonicities with their
corresponding experimental values supports the quality of the resultant
potential energy curves.\bigskip{}

\end{abstract}
\maketitle
In new quantum Monte Carlo calculations (in course of publication),
Amovilli and March have studied N$_{2}$ and N$_{2}^{+}$. This work
has motivated us to consider P$_{2}$ and P$_{2}^{+}$, but now using
the one-particle reduced density matrix (1-RDM) functional theory
in its spectral representation, i.e., the natural orbital functional
(NOF) theory \cite{Piris2007}. As a result, the ground state energy
is expressed in terms of natural orbitals and their occupation numbers. 

Recent advances \cite{Piris2013a,Pernal2013,Piris2013e} have led
to the formulation of the first strictly N-representable functional
known under the acronym PNOF5 \cite{Piris2011}. The latter belongs
to a series of functionals \cite{Piris2013b} based on a reconstruction
of the two-particle reduced density matrix (2-RDM) in terms of the
1-RDM by ensuring necessary N-representability positivity conditions
on the 2-RDM \cite{Piris2006}. This functional has proved to give
a remarkably accurate description of systems with near-degenerate
one-particle states and of homolytic dissociation processes \cite{Matxain2011,Lopez2012,Ruiperez2013,Piris2014d}.
PNOF5 provides a very appealing one-electron picture \cite{Matxain2012a,Matxain2012,Piris2013,Matxain2013a},
and has been successfully used \cite{Piris2012a,Piris2015,Piris2015b}
to predict vertical ionization potentials by means of the extended
Koopmans\textquoteright{} theorem (EKT) \cite{Day1974,Day1975,Smith1975,Morrell1975}.

PNOF5 is an orbital-pairing approach that involves coupling each orbital
$g$, below the Fermi level $(g\leq N/2)$, with $N_{g}$ orbitals
above it, which is reflected in the sum rule for the occupation numbers,
namely,
\begin{equation}
\sum_{p\in\Omega_{g}}n_{p}=1;\quad g=\overline{1,\nicefrac{N}{2}}\label{sumrule_n}
\end{equation}
In eqn. (\ref{sumrule_n}), $\Omega_{g}$ is the subspace containing
the orbital $g$ and $N_{g}$ coupled orbitals to it. Moreover, we
consider that these subspaces are mutually disjoint $\left(\Omega_{g1}\cap\Omega_{g2}=\textrm{Ø}\right)$,
i.e., each orbital belongs only to one subspace $\Omega_{g}$. The
PNOF5 energy for a singlet state of an $N$-electron system can be
cast as
\begin{equation}
E={\displaystyle \sum\limits _{g=1}^{\nicefrac{N}{2}}}\mathrm{\:}E_{g}+\sum\limits _{f\neq g}^{\nicefrac{N}{2}}\sum\limits _{p\in\Omega_{f}}\sum\limits _{q\in\Omega_{g}}n_{p}n_{q}\left(2\mathcal{J}_{qp}-\mathcal{K}_{qp}\right)\label{PNOF5e}
\end{equation}
\begin{equation}
E_{g}=\sum\limits _{p\in\Omega_{g}}n_{p}\left(2\mathcal{H}_{pp}+\mathcal{J}_{pp}\right)+\sum\limits _{p,q\in\Omega_{g},p\neq q}\Pi\left(n_{p},n_{q}\right)\mathcal{L}_{qp}\label{Epair}
\end{equation}
\begin{equation}
\quad\Pi\left(n_{p},n_{q}\right)=\left\{ \begin{array}{cc}
-\sqrt{n_{p}n_{q}}\,, & _{p=g\: or\:}{}_{q=g}\\
\;\sqrt{n_{p}n_{q}}\,, & _{p,q>\nicefrac{N}{2}}
\end{array}\right.\label{Pipq}
\end{equation}

where $\mathcal{H}_{pp}$ denotes the one-particle matrix elements
of the core-Hamiltonian. $\mathcal{J}_{pq}=\left\langle pq|pq\right\rangle $
and $\mathcal{K}_{pq}=\left\langle pq|qp\right\rangle $ are the usual
direct and exchange integrals, respectively. $\mathcal{L}_{pq}=\left\langle pp|qq\right\rangle $
is the exchange and time-inversion integral \cite{Piris1999}. In
this work, we consider only one orbital coupled to $g$ in each subspace,
the so-called perfect pairing.

\begin{figure}[t]
\caption{\label{PEC for P2}Potential energy curve of P$_{2}$ obtained at
the PNOF5/cc-pVTZ level of theory. \vspace{1cm}
}

\noindent \centering{}\includegraphics[scale=0.3]{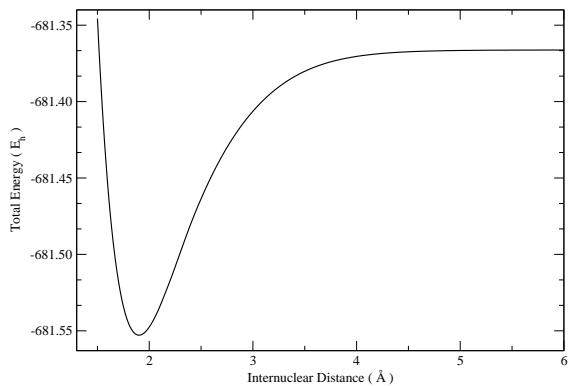}
\end{figure}

Fig. \ref{PEC for P2} shows the calculated values in Hartrees ($E_{h}$)
of the total ground-state energy as a function of the internuclear
distance $R$, in $\textrm{\AA}$, for the neutral Phosphorous diatomic
molecule using the correlation-consistent valence triple-$\zeta$
basis set (cc-pVTZ) developed by Woon and Dunning \cite{Phos-cc-pVTZ}.
The dissociation limit corresponds to a two-fold degeneracy with the
generation of two doublet atomic states. We note that the equilibrium
bond length is given by PNOF5 as $R_{e}=1.901$ \AA{}, which compares
quite well with the experimental value of $1.893$ \AA{} \cite{NIST}.
Besides, we have obtained an excellent agreement for the dissociation
energy (D$_{e}$). Indeed, the calculated PNOF5 D$_{e}$ is 117.1
kcal/mol, whereas the experimental mark (118.0 kcal/mol) taken from
a combination of Refs. \cite{NIST} and \cite{Luo2007}, differs only
in 0.9 kcal/mol. It is also remarkable that PNOF5 is able to reproduce
the correct integer number of electrons on the dissociated atoms.

We turn next to the NOF results obtained for the positive molecular
ion P$_{2}^{+}$. Our approach for the computation of the energy of
the cation as a function of the internuclear distance $R$, consists
of adding the vertical ionization potential (IP) to the previously
obtained energy for the neutral dimer at the corresponding $R$. The
vertical electron detachment energy is computed from the EKT formalism.
The equation for the EKT may be derived by expressing the wavefunction
of the ($N-1$)-electron system as the following linear combination
\begin{equation}
\left\vert \Psi^{N-1}\right\rangle =\sum\limits _{i}C_{i}\widehat{a}_{i}\left\vert \Psi^{N}\right\rangle \label{psim1}
\end{equation}

In Eq. (\ref{psim1}), $\widehat{a}_{i}$ is the annihilation operator
for an electron in the spin-orbital $\left\vert \phi_{i}\right\rangle =$$\left\vert \varphi_{p}\right\rangle \otimes$$\left\vert \sigma\right\rangle \,\left(\sigma=\alpha,\beta\right)$,
$\left\vert \Psi^{N}\right\rangle $ is the wavefunction of the $N$-electron
system, $\left\vert \Psi^{N-1}\right\rangle $ is the wavefunction
of the ($N-1$)-electron system, and $\left\{ C_{i}\right\} $ are
the set of coefficients to be determined. Optimizing the energy of
the state $\Psi^{N-1}$ with respect to the parameters $\left\{ C_{i}\right\} $
and subtracting the energy of $\Psi^{N}$, gives the EKT equations
as a generalized eigenvalue problem,
\begin{equation}
\mathbf{FC}=\mathbf{\Gamma C\mathbf{\nu}}\label{ekt}
\end{equation}

where $\mathbf{\nu}$\ are the EKT IPs and the metric matrix $\mathbf{\Gamma}$
is the 1-RDM with the occupation numbers along the diagonal and zeros
in off-diagonal elements. The transition matrix elements are given
by
\begin{equation}
F_{ji}=\left\langle \Psi^{N}\right\vert \widehat{a}_{j}^{\dagger}\left[\widehat{\mathcal{H}},\widehat{a}_{i}\right]\left\vert \Psi^{N}\right\rangle \label{fv}
\end{equation}

\begin{figure}[t]
\caption{\label{PEC for P2+}Potential energy curve of P$_{2}^{+}$ obtained
at the PNOF5(P$_{2}$)+EKT/cc-pVTZ level of theory. \vspace{1cm}
}

\noindent \centering{}\includegraphics[scale=0.3]{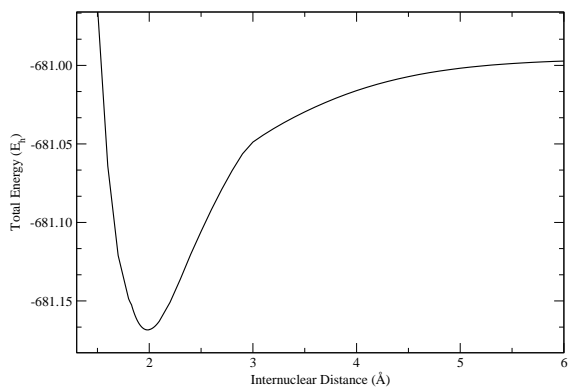}
\end{figure}

Considering a spin-restricted theory, it is not difficult to obtain
the transition matrix elements $F_{qp}=-\lambda_{qp}$, being $\left\{ \lambda_{qp}\right\} $
the set of symmetric Lagrange multipliers associated with the orthonormality
of natural orbitals. Equation (\ref{ekt}) can be transformed by a
canonical orthonormalization using $\mathbf{\Gamma}^{-1/2}$. Hence,
the diagonalization of the matrix $\mathbf{\nu}$ with the elements
\begin{equation}
\nu_{qp}=-\frac{\lambda_{qp}}{\sqrt{n_{q}n_{p}}}\label{niu}
\end{equation}
yields IPs as eigenvalues.

One test of the reliability of this procedure constitutes the prediction
of the first IPs at the equilibrium and dissociation limit, which
correspond to the dimer and atom species, respectively. PNOF5-EKT
predicts a value of 10.57 eV for P$_{2}$, which reproduces satisfactorily
the experimental vertical ionization energy of 10.62 eV, whereas the
calculated IP for P atom is 10.08 eV, which underestimates in 0.4
eV the experimental mark \cite{NIST}. In the case of P$_{2}$, the
experimental adiabatic ionization energy is 10.53 eV \cite{NIST}.

The potential energy curve obtained for the cation P$_{2}^{+}$ using
the cc-pVTZ is shown in Fig. \ref{PEC for P2+}. Note that the proposed
E(P$_{2}$)+EKT methodology affords a proper dissociation curve. The
equilibrium bond length $R_{e}$ is found to change significantly
from 1.901 \AA{} for neutral P$_{2}$ to 1.985 \AA{} for the positive
ion, in excellent agreement with the experimental value of 1.986 \AA{}
\cite{NIST}. On the other hand, the E(P$_{2}$)+EKT calculations
slightly underestimate the dissociation energy for P$_{2}^{+}$ in
7.6 kcal/mol with respect to the experimental value of 115.9 kcal/mol
\cite{NIST,Luo2007}. This result agrees with the mentioned above
underestimation of the IP for atomic species obtained using the EKT
at the dissociated neutral dimer.

\begin{table*}[t]
\caption{\label{tab:Comparison}Comparison with the experimental values of
the equilibrium bond length (R$_{e}$, in $\textrm{\AA}$), dissociation
energy (D$_{e}$, in kcal/mol), harmonic vibrational frequency ($\omega_{e}$,
in cm$^{-1}$), first-order ($\omega_{e}x_{e}$ , in cm$^{-1}$) and
second-order ($\omega_{e}y_{e}$ , in cm$^{-1}$) anharmonicity constants.
Properties were calculated at the PNOF5/cc-pVTZ level of theory. \medskip{}
}

\noindent \centering{}%
\begin{tabular}{c|ccccc|ccccc}
\hline 
 & \multicolumn{5}{c|}{PNOF5} & \multicolumn{5}{c}{Experiment \cite{NIST,Luo2007}}\tabularnewline
\hline 
\hline 
 & R$_{e}$ & D$_{e}$ & $\omega_{e}$ & $\omega_{e}x_{e}$ & $\omega_{e}y_{e}$ & R$_{e}$ & D$_{e}$ & $\omega_{e}$ & $\omega_{e}x_{e}$ & $\omega_{e}y_{e}$\tabularnewline
\hline 
P$_{2}$ & 1.901 & 117.1 & 783.3 & 2.912 & -0.0178 & 1.893 & 118.0 & 780.8 & 2.835 & -0.0046\tabularnewline
P$_{2}^{+}$ & 1.985 & 108.3 & 692.8 & -3.951 & 0.8423 & 1.986 & 115.9 & 672.2 & 2.740 & -\tabularnewline
\hline 
\end{tabular}
\end{table*}

In Table \ref{tab:Comparison}, selected electronic properties, including
bond lengths, dissociation energies, harmonic vibrational frequencies
and the anharmonicity constants can be found for both potential energy
curves. The good quality of the resultant potential energy curve for
P$_{2}$ is illustrated by the nice agreement with the experimental
marks of the harmonic vibrational frequency and the anharmonicities.
Conversely, PNOF5+EKT overestimates the harmonic vibrational frequency
for the cation in 22.6 cm$^{-1}$. 

In P$_{2}^{+}$, the obtained value for the anharmonicity constant
($\omega_{e}x_{e}$) deserves special attention. In fact, there is
a discontinuity in the derivative of the potential energy with respect
to the internuclear distance around 3 $\textrm{\AA}$. This behavior
is due to the change in the solutions passing from the equilibrium
region to the dissociated molecule in the neutral species, which leads
finally to a wrong value of $\omega_{e}x_{e}$ in the calculated potential
energy curve for the ion through the EKT.

In summary, the calculation of potential energy curves for P$_{2}$
and P$_{2}^{+}$ has been achieved solely from an appropriate 2-RDM
that leads to a strictly N-representable NOF for the neutral dimer.
The agreement with accessible experimental data is found to be quite
satisfactory without recourse to relevant wavefunctions.

\selectlanguage{american}%

\section*{Acknowledgements}

Financial support comes from Eusko Jaurlaritza (Ref. IT588-13) and
Ministerio de Economía y Competitividad (Ref. CTQ2015-67608-P). The
authors thank for technical and human support provided by IZO-SGI
SGIker of UPV/EHU and European funding (ERDF and ESF). NHM completed
his contribution to this article during a stay at DIPC. He thanks
Professor P. M. Echenique and Dr. A. Ayuela for very generous support.

\selectlanguage{english}%
\smallskip{}


\end{document}